\documentclass[cits]{PoS}

\def\bg#1{\mbox{\boldmath$#1$}}

\newcommand{\beq}{\begin{eqnarray}}
\newcommand{\eeq}{\end{eqnarray}}
\newcommand{\be}{\begin{eqnarray*}}
\newcommand{\ee}{\end{eqnarray*}}

\newcommand{\D}{{\cal D}}
\newcommand{\Pom}{{\hspace{ -0.1em}I\hspace{-0.25em}P}}
\newcommand{\vphot}{{\gamma^*}}

\newcommand{\QQ}{{\scriptscriptstyle{{Q \bar{Q}}}}}

\def\lsim{\raise0.3ex\hbox{$<$\kern-0.75em\raise-1.1ex\hbox{$\sim$}}}
\def\gsim{\raise0.3ex\hbox{$>$\kern-0.75em\raise-1.1ex\hbox{$\sim$}}}

\title{Energy dependence of nuclear effects in hadron-nucleus collisions}

\ShortTitle{Energy dependence of nuclear effects in hadron-nucleus collisions}

\author{\speaker{K.~Tywoniuk}, I.~C.~Arsene, L.~Bravina and
  E.~Zabrodin\\ 
        University of Oslo\\
        E-mail: \email{konrad.tywoniuk@fys.uio.no}\\}

\author{A.~B.~Kaidalov\\
        ITEP, Moscow}

\abstract{The energy dependence of light and heavy particle production
  in hadron-nucleus collisions
  is discussed. Whereas the production mechanism at lower energies can
  be understood in the Glauber rescattering picture, experimental data
  at RHIC indicate that particles are mostly produced in coherent
  processes. The importance of energy-momentum conservation is shown
  to be crucial at forward rapidities for the whole energy range. We
  also discuss the behaviour of $\alpha (x_F)$ with energy for light
  particles and $J/\psi$. Finally, we make predictions for the future
  LHC experiment.
}

\FullConference{High-pT physics at LHC\\
		 March 23-27 2007\\
		 University of Jyv\"askyl\"a, Jyv\"askyl\"a, Finland}

\begin{document}

\section{Introduction}
\begin{figure}[b!]
  \begin{minipage}{0.5\linewidth}
    \centering
    \includegraphics[width=.5\textwidth,height=2.5cm]{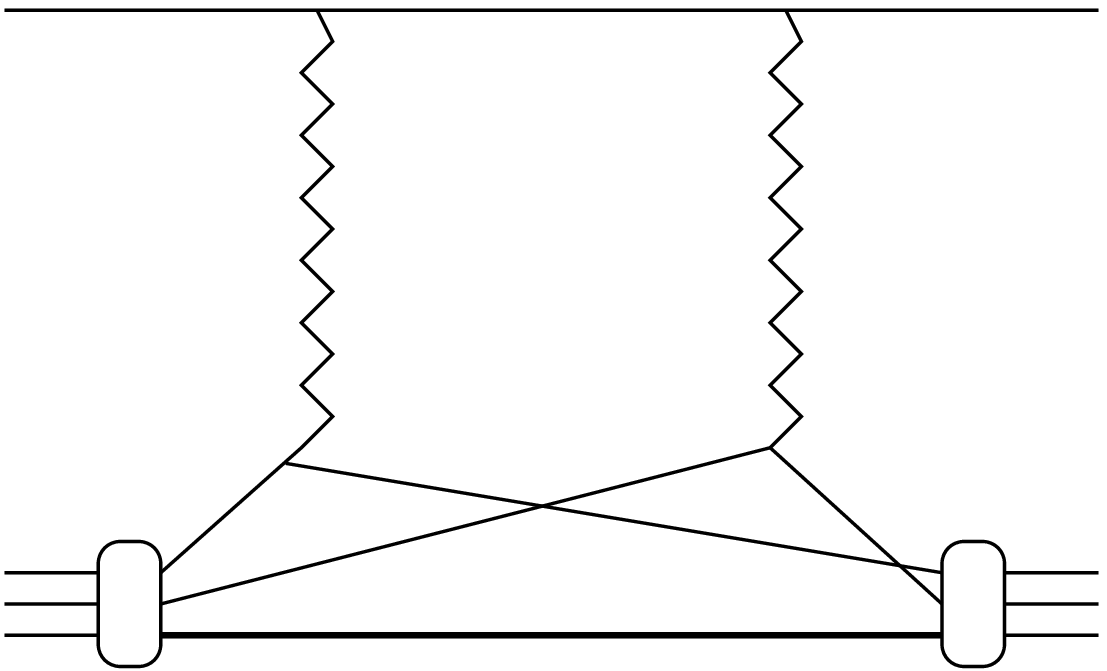}
  \end{minipage}
  \begin{minipage}{0.5\linewidth}
    \centering
    \includegraphics[width=0.5\textwidth,height=2.5cm]{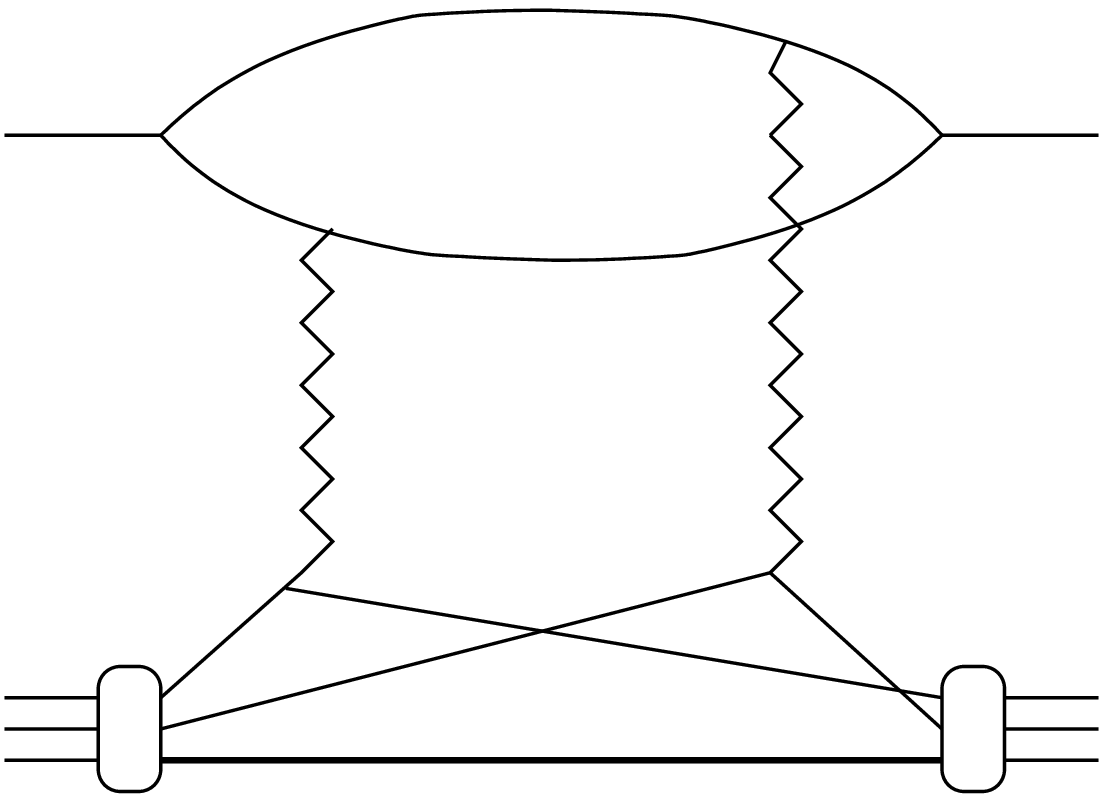}
  \end{minipage}
  \caption{Planar (left) and non-planar (right) diagram of multiple
    scattering. The former decreases as $\sim 1/E$ above the critical
    energy, the latter is controlled by $t_{min}$ effects.}
  \label{fig:PlanarDiagram}
\end{figure}
In addition to being an important tool for modeling nucleus-nucleus
collisions (AA), proton-nucleus (pA) collisions are also interesting
by themselves as they probe nuclear effects related to multiple
scattering and possible hadronization in nuclear matter. The
origin of these effects is still under debate. In recent years most of
theoretical activity has focused on novel high-energy
nuclear effects, such as parton saturation, yet it seems that past and
present energies (RHIC) do not provide enough phase space for these
effects to appear. It will be our task in this paper to review some
well-known low-energy models and to introduce high-energy effects
(shadowing) to describe data on light and heavy particles in pA
collisions in the energy range $\sqrt{s} = 17.3 - 5500$ GeV.

\begin{figure}[t!]
  \centering
  \includegraphics[width=0.5\linewidth]{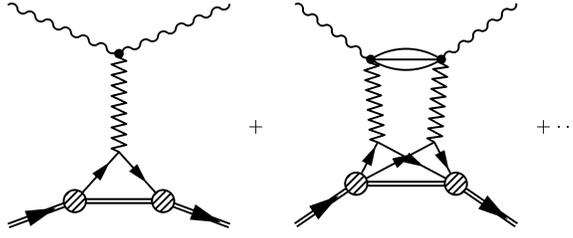}
  \caption{Glauber-Gribov series: the single and double scattering
    contribution to the total $\vphot N$ cross section.}
  \label{fig:diag1}
\end{figure}

Our starting point is noticing that a significant change in the
underlying dynamics of a hadron-nucleus
collision takes place with growing energy of the incoming particles.
At low energies, the total cross section is well described within
the probabilistic Glauber model \cite{Glauber59}. In the reggeon
approach these interactions are
described by so-called planar diagrams
depicted in Fig.~\ref{fig:PlanarDiagram} (left).
At higher energies, $E > E_{\scriptstyle{C}} \sim
m_{\scriptstyle{N}} \mu R_A$ ($\mu$ is a characteristic hadronic scale,
$\mu \sim 1$ GeV, and $R_A$ is the radius of the nucleus)
corresponding to a coherence length
\beq 
\label{eq:cohlenght} 
l_{\scriptscriptstyle{C}} \;=\; \frac{1}{2 \, m_N \, x} \;, 
\eeq
the typical hadronic fluctuation length can become of the order of,
or even bigger than, the nuclear radius and there will be
coherent interaction of hadron constituents with several
nucleons of the nucleus. In this energy range, the contribution from planar
diagrams is damped by a factor $\sim 1/E$ \cite{Mandelstam63} and the dominant
contribution arises from non-planar diagrams, shown in Fig.~\ref{fig:PlanarDiagram}
(right).
The sum of all diagrams was calculated by
Gribov \cite{Gribov69,Gribov70}, who introduced a correction to the
Glauber series by taking into account the diffractive intermediate
states. The forward hadron-nucleus ($hA$ or $\vphot A$)
scattering amplitude can then be written as the sum of diagrams shown in
Fig.~\ref{fig:diag1}, i.e. as 
\beq
\sigma_{\vphot A} \;=\; A\sigma_{\vphot N} \,+\, \sigma_{\vphot A}^{(2)} \,+\, ...\;,
\label{eq:sum}
\eeq
In Eq.~(\ref{eq:sum}), the first term simply represents the Glauber 
elastic contribution and subsequent terms describe multiple
interactions of the incoming probe with the nucleons in the target nucleus.
The space-time picture analogy to the
Glauber series is however lost, as the interactions with
different nucleons of the nucleus occur nearly simultaneous in time.
This phenomenon is related to inelastic shadowing corrections.

An additional effect which comes into play at high energies, is the
possibility of interactions between soft partons of the different
nucleons in the nucleus. In the Glauber-Gribov theory this
corresponds to interactions between Pomerons.
The necessity to include such ``enhanced'' diagrams at high
energies can be related to unitarization of all amplitudes.
In particular, diagrams involving triple-Pomeron interactions are
related to large-mass intermediate states in Fig.~\ref{fig:diag1}
which give a dominant contribution to shadowing.
There is a connection between these effects and saturation effects in
the parton picture.\par
\begin{figure}[t!]
  \begin{minipage}{0.5\linewidth}
    \centering
    \includegraphics[width=.9\textwidth]{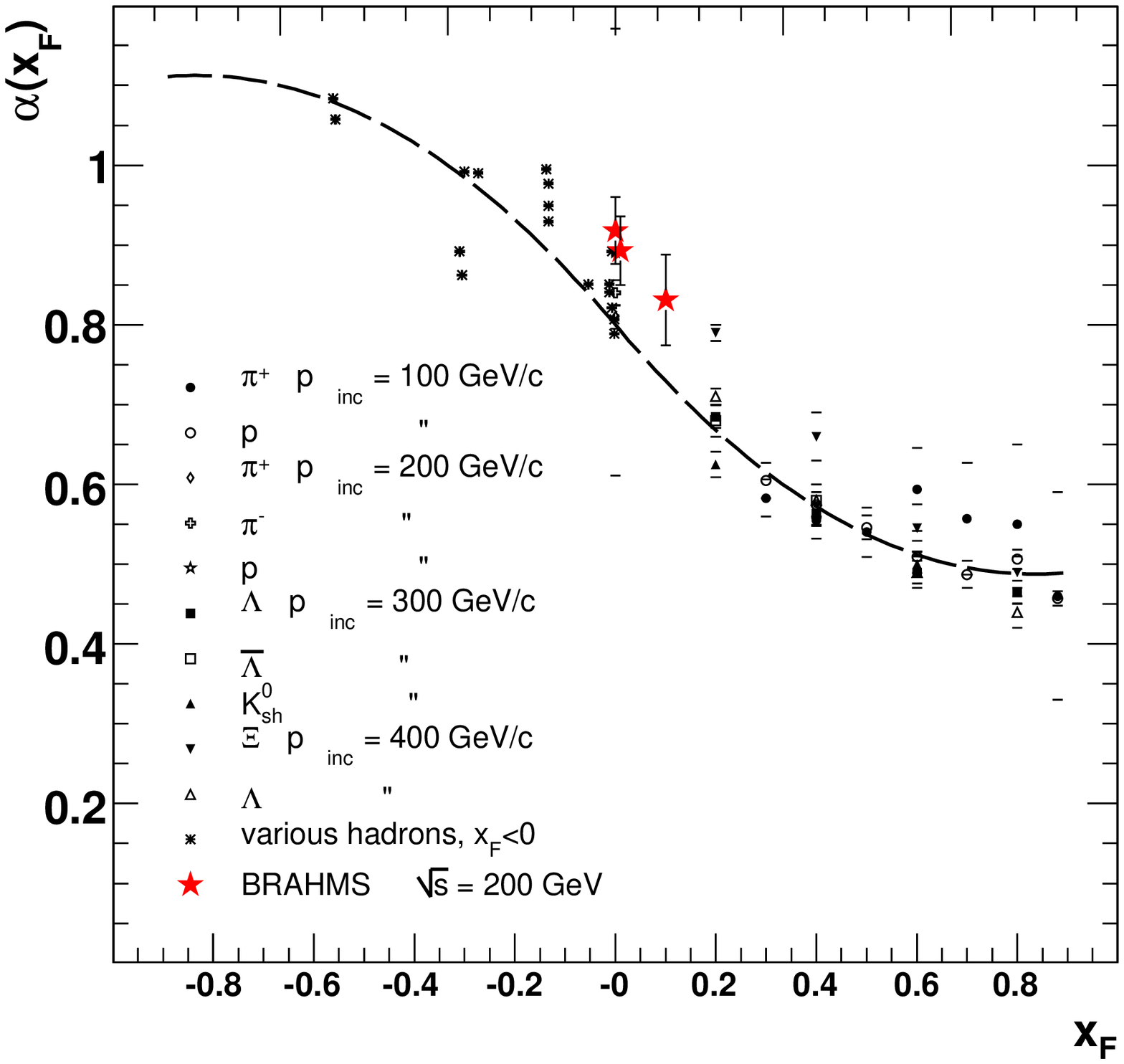}
  \end{minipage}
  \begin{minipage}{0.5\linewidth}
    \centering
    \includegraphics[width=0.9\textwidth]{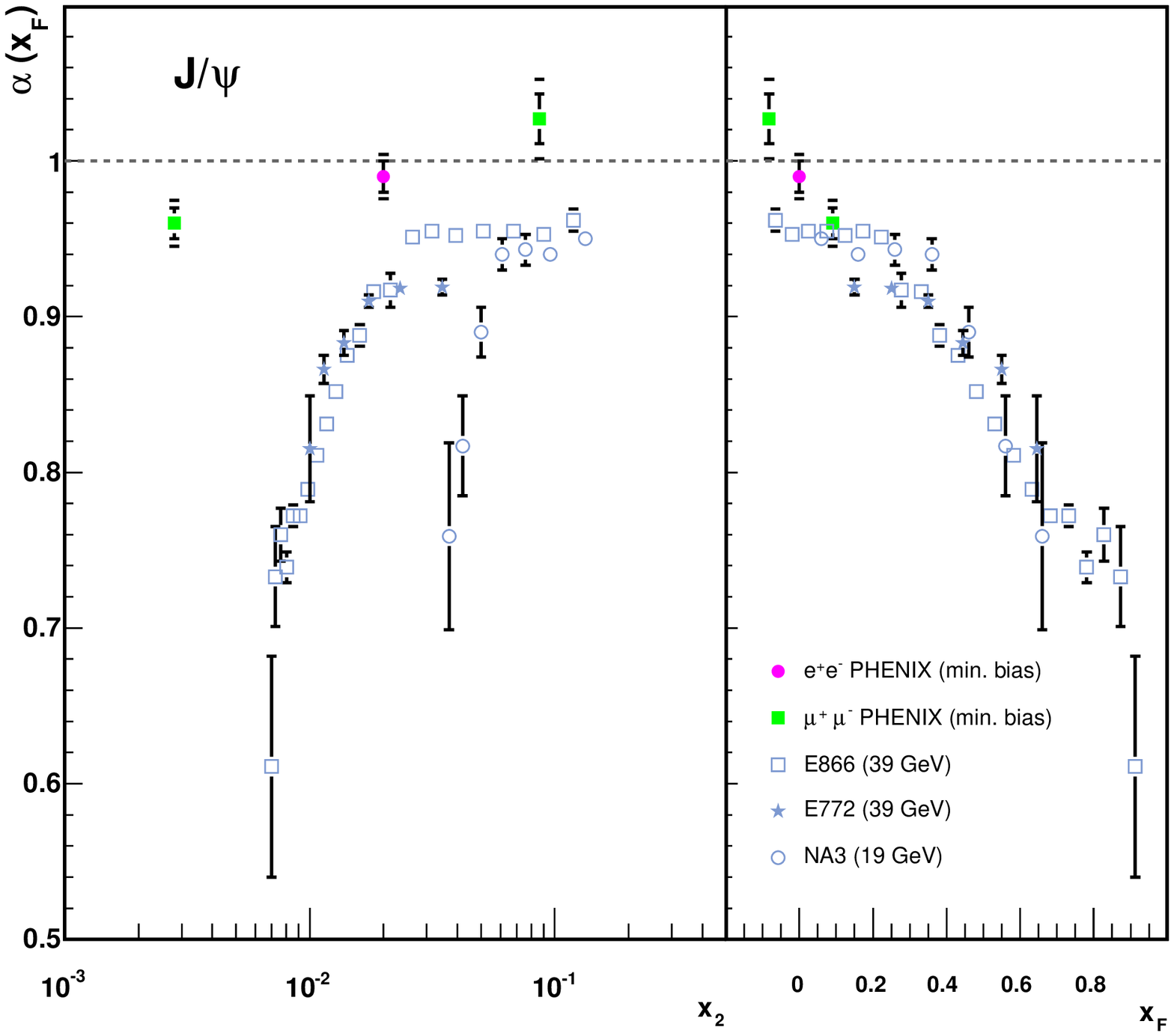}
  \end{minipage}
  \caption{$\alpha$ vs. $x_F$ (and $x_2$) in proton-nucleus collisions at
    different energies for production of light particles (left) and
    $J/\psi$ (right). Experimental
    data are taken from \cite{Alde91,Leitch00,Shahoyan02,Arsene04,Adler03,Geist91}.}
  \label{fig:AlphaExperiment}
\end{figure}

In pA collisions, nuclear effects are usually discussed in terms of
the power-law parameterization
\beq
\label{eq:param}
\frac{\mbox{d}\sigma_{pA}}{\mbox{d}^3p} \;=\;
\frac{\mbox{d}\sigma_{pN}}{\mbox{d}^3 p} \, A^{\alpha(x_{\scriptscriptstyle{F}})} \;,
\eeq
where $\sigma_{pA}$ ($\sigma_{pN}$) is the inclusive cross
section off a nucleus (nucleon). The function $\alpha(x_F)$
characterizes nuclear effects at different longitudinal momentum
fractions of the produced particle. For a large range of
energies, $\alpha$ exhibits a very interesting scaling in $x_F$ for
both light and heavy particle production as seen in
Fig.~\ref{fig:AlphaExperiment}. For $J/\psi$,
$\alpha (x_F)$ 
decreases from 0.95 at $x_F \approx 0$ to values $\alpha \sim 0.75$ at $x_F
\simeq 0.8$ thus indicating an increase of absorption as $x_F$
increases \cite{Alde91,Leitch00,Shahoyan02}. No scaling in Bjorken $x$
of the nucleus, $x_2$, is
observed, indicating breaking of QCD factorization at these energies.
The suppression of light particles follow a similar trend.
Another striking feature is the rather large suppression of low-energy
data at $x_F = 0$. These features are reproduced by models invoking
mechanisms of attenuation and/or energy-loss in nuclear matter
\cite{Kopeliovich05,Kopeliovich01,Vogt00,Boreskov93}.

Recent data from RHIC experiment on charged hadron \cite{Arsene04} and
$J/\psi$ \cite{Adler03} production in dAu collisions at $\sqrt{s} =
200$ GeV are also shown in
Fig.~\ref{fig:AlphaExperiment} (data from \cite{Arsene04} have been
integrated over $p_\bot$ between 0.7-1.0 GeV/c). The
suppression at $x_F = 0$ is smaller compared to lower energies
contrary to the expectation of many theoretical models. Despite the
limited kinematics accessible in collider experiments, the RHIC data
also hints towards a breaking of $x_F$ scaling. 

The behavior of $\alpha(x_F)$ with energy allows a natural
explanation within the Gribov theory of multiparticle production
described above. At very high energies Abramovsky-Gribov-Kancheli
(AGK) cutting rules \cite{Abramovsky74} lead to cancellation of the
Glauber-type diagrams in the central rapidity region and only enhanced
diagrams, or in other words shadowing of small-$x$ partons, contribute
to $\alpha < 1$. Experimental data indicate, that the transition from
incoherent to coherent particle production happens at RHIC energies
for $J/\psi$ production.
In the following we will discuss these trends in more detail and
present calculations involving gluon shadowing and effects of
energy-momentum conservation.

\section{Gribov inelastic shadowing}

In the relativistic Gribov theory \cite{Gribov69,Gribov70} the
collision proceeds through simultaneous interactions of the projectile
with nucleons in the nucleus and therefore the intermediate
states, shown in Fig.~\ref{fig:diag1}, are no longer the same as the
initial state.
The multiparticle content of the diagrams is given by
AGK cutting rules \cite{Abramovsky74}, where the intermediate states are
on-shell. The cut contribution of the double rescattering diagram can be
expressed in terms of diffractive
deep inelastic scattering (DDIS). 
The variable $\beta =
\frac{Q^2}{Q^2+M^2} = x /x_\Pom$ plays the same role for the Pomeron
as the Bjorken variable, $x$, for the nucleon. 
We assume that the 
amplitude of the process is purely imaginary. The contribution from
the second term in Eq.~(\ref{eq:sum}) to
the total $\vphot A$ cross section is given by
\beq
\sigma^{(2)}_{\vphot A} \;=\;& & \; -4\pi A(A-1)\,\int \mbox{d}^2b \,
\int_{M^2_{min}}^{M^2_{max}} \mbox{d}M^2 \, \left[ \frac{\mbox{d}
    \sigma^{\D}_{\vphot \scriptscriptstyle{N}} (Q^2, x_\Pom, \beta)}{\mbox{d} M^2 \,
    \mbox{d}t}\right]_{t=0} \, |F_A(q_L,b)|^2 \;,
\label{eq:ExactFormula}
\eeq
where 
\beq
|F(q_L,b)|^2 \;=\; \int_{-\infty}^\infty \mbox{d}z \, \rho_A(b,z) \, e^{i
  q_{\scriptstyle{L}}z} \;,
\label{eq:FormFactKK}
\eeq
is the so-called longitudinal form factor and $q_L = \sqrt{-t_{min}} =
m_N x_\Pom$ \cite{Karmanov73}. 

In Eq.~(\ref{eq:ExactFormula}), $M^2_{min}$ corresponds to
the minimal mass of the diffractively 
produced hadronic system, $M^2_{min} = 4m_{\pi}^2 = 0.08 \mbox{
  GeV}^2$, and $M^2_{max}$ is chosen according to the
condition: $x_\Pom \le x_\Pom^{max}$.
We use the standard choice for $x_\Pom^{max} = 0.1$
\cite{Kaidalov79}. It is convenient as it guarantees the disappearance of
nuclear shadowing at $x \sim 0.1$ in accord to experimental data.

We are interested in calculating shadowing for small-$x$ quarks and
gluons and consider because of this DDIS on nucleons where this
information can be extracted.
Note that since Eq.~(\ref{eq:ExactFormula}) is
obtained under very general assumptions, i.~e. analyticity and
unitarity, it can be applied for arbitrary values of $Q^2$ provided
$x$ is very small \cite{Brodsky02}.

\subsection{Diffractive gluon distribution}
The cross sections of diffractive processes are
expressed through structure functions, which are in turn
associated with distribution functions of partons in the Pomeron. 
The relation between 
the diffractive cross section and the diffractive structure function
is given by
\be
\left[ \frac{\mbox{d}
    \sigma^{\D}_{\vphot N} (Q^2, x_\Pom, \beta)}{\mbox{d} M^2 \,
    \mbox{d}t} \right]_{t=0} \;=\; \frac{4\pi^2 \alpha_{em} \, B}{Q^2\,
  (Q^2 + M^2)} \, x_\Pom \, F_{2\D}^{(3)} (Q^2, x_\Pom, \beta) \,
\ee
Assuming Regge factorization we write the
diffractive structure function as
\beq
\label{eq:reggefact}
F_{2\D}^{(3)}(x_\Pom, Q^2, \beta) \;=\; f_\Pom(x_\Pom) \, F(\beta,Q^2) \;,
\eeq
where the first factor is referred to as the ($t$-integrated) Pomeron
flux and the second factor,
$F(\beta,Q^2)$, is the Pomeron structure function. 

Until recently, the diffractive structure function has been poorly
known. In particular, the diffractive gluon density was affected by
large uncertainty since it is not measured directly in experiment.
Results of new high-precision measurements of
the diffractive parton distribution functions (DPDFs) presented 
by the H1 Collaboration \cite{H1new1,H1new2} give important
constraints to our model of shadowing. 
Due to the indirect extraction of the diffractive gluon
density, $\beta g^\D(\beta,Q^2)$, from experimental data, two fits were
presented, FIT A and FIT 
B, reflecting the systematic uncertainty of the procedure.
Furthermore, a combined fit to DDIS data and 
diffractive di-jets \cite{Mozer06} results in a curve similar
to FIT B yet with a slightly smaller gluon density. For completeness,
we also compare the new results with the old H1 parameterization
\cite{H1abs} presented in 2002. For details on extracted DPDFs and
corresponding Pomeron parameters, $\alpha_\Pom (0)$ and
$\alpha_\Pom'$, we refer the reader to the experimental papers
\cite{H1new1,H1new2}.

The gluon distribution is
almost a factor of 10 bigger than the quark distribution at the same $Q^2$
for a broad region of $\beta$ \cite{H1new1,H1new2}. 
Therefore, in the relevant kinematical range for
hadron-nucleus and nucleus-nucleus collisions at RHIC and
LHC, the gluon density dominates. In what follows, we 
will therefore consider structure functions of gluons in nuclei. The
first term in Eq.~(\ref{eq:sum}) is then proportional to $A G_N(x,Q^2)$,
while the second rescattering term is correspondingly equal to
$-A(A-1)G_N(x,Q^2) \, \int\mbox{d}^2b \, T^2_A(b) \,f(x,Q^2)$, where
\beq
\label{eq:fgluon}
f \left(x, Q^2 \right) \;=\; 4\pi \, \int^{x_\Pom^{max}}_x
\mbox{d}x_\Pom \, B(x_\Pom) f_\Pom(x_\Pom) \, \frac{ \beta
  g^\D (\beta,Q^2)}{G_N (x,Q^2)} \, F_A^2 (t_{min}) \;,
\eeq
with $B(x_\Pom)$ being the diffractive slope parameter.
The gluon distribution of the nucleon, $G_N(x,Q^2) = xg(x,Q^2)$, was
taken from the CTEQ6M parameterization \cite{Pumplin02}. The ratio
under the integral in Eq.~(\ref{eq:fgluon}) is hence the ratio of
gluon density in the Pomeron and in the nucleon.

\subsection{Models of multiple scattering}\label{sec:shad}
\begin{figure}
  \centering
  \includegraphics[width=0.55\textwidth]{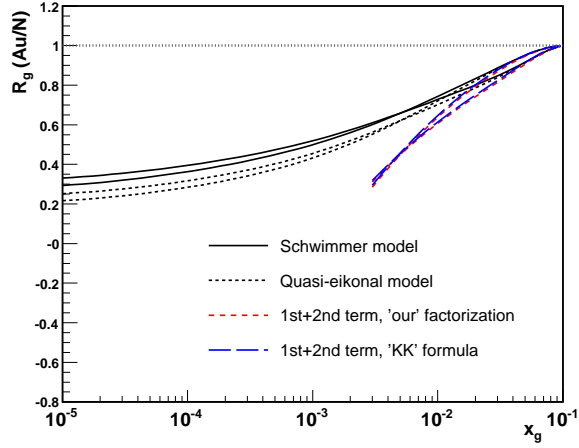}
  \caption{Comparison of different models for higher order
    rescatterings. See text for details.}
  \label{fig:ModelComp}
\end{figure}
The summation of all rescatterings in Eq.~(\ref{eq:sum}) is model
dependent.
In writing Eq.~(\ref{eq:fgluon}) we have assumed that the following
factorization holds
\beq
|F(q_L,b)|^2 \;=\; T^2_A(b) \, F^2_A\left(t_{min} \right) \;, 
\label{eq:FFfactorization}
\eeq
where $T_A (b) = \int^{+\infty}_{-\infty}
\mbox{d}z \, \rho_A ({\bg 
  b}, z)$ is the nuclear density profile and $F_A$ is given by 
\beq
\label{eq:FA}
F_A (t_{min}) \;=\; \int \mbox{d}^2b \, J_0 (\sqrt{-t_{min}}b) \, T_A
(b) \;.
\eeq
Here $J_0(x)$ denotes the Bessel function of the first kind. 
Equation~(\ref{eq:FFfactorization}) is an identity for nuclear
densities which depend separately on $b$ and $z$, 
however we have checked that calculations with a Woods-Saxon nuclear
density profile lead to negligible corrections to the exact
expression. For consistency, we show calculations
of $(\sigma^{(1)} +
\sigma^{(2)})/\sigma_{\vphot N}$ in Fig.~\ref{fig:ModelComp} where for the dash-double-dotted
curve Eq.~(\ref{eq:FFfactorization}) (denoted
'our' factorization) has been employed and the 
long-dashed curve shows the result using directly
Eq.~(\ref{eq:ExactFormula}) (denoted 'KK' formula) \cite{Karmanov73}. The two
curves practically coincide and indicate furthermore that higher order corrections are essential.

We will now compute the total $\vphot A$ cross section using two
models for higher-order rescatterings: a Schwimmer 
unitarization \cite{Schwimmer75} which is obtained from a summation of
fan-diagrams with triple-Pomeron interactions and a quasi-eikonal
unitarization. Nuclear shadowing is studied in terms of the ratios of
cross sections per nucleon for different nuclei, defined as
\beq
\label{eq:ratio}
R (A/B) \;=\; \frac{B}{A} \, \frac{\sigma_{\vphot A}}{\sigma_{\vphot B}} \;,
\eeq
as a function of $x$, which can in turn be expressed via structure
functions of the different nuclei. The simplest case is B$=$N, then
\beq
\label{eq:sch}
R_g^{Sch} \left( A/N \right) (x) \;&=&\; \int \mbox{d}^2b \, \frac{T_A
  (b)}{1 \,+\, (A-1) f(x, Q^2) T_A (b) } \\
\label{eq:eik}
R_g^{eik} \left( A/N \right) (x) \;&=&\; \int \mbox{d}^2b \, \frac{1}{2(A-1)
  f(x,Q^2)} \, \left\{ 1 \,-\, \exp \left[- 2 (A-1) T_A (b) \, f(x,Q^2)
  \right] \right\}
\eeq
for the Schwimmer and quasi-eikonal models, respectively. In
Eqs.~(\ref{eq:sch}) and (\ref{eq:eik}), $f(x,Q^2)$ is given by
Eq.~(\ref{eq:fgluon}). Both expressions (\ref{eq:sch}) and
(\ref{eq:eik}), expanded to the first non-trivial order, reproduce the
the second order rescattering result in Eq.~(\ref{eq:ExactFormula}). 
The gluon shadowing ratio $R_g$ for $x <
0.1$ is shown in Fig.~\ref{fig:ModelComp} where the calculations for
Schwimmer (quasi-eikonal) model are depicted with a solid (dotted)
curve. 

The quasi-eikonal model gives a stronger shadowing effect than the
Schwimmer model. In our framework shadowing can also be studied as a
function of the impact parameter $b$, given by
\beq
\label{eq:bshad}
R_g^{Sch} \left( A/N \right) (b) \;&=&\; \frac{1}{1 \,+\, (A-1) f(x,
  Q^2) T_A (b) } \;, \\
R_g^{eik} \left( A/N \right) (b) \;&=&\; \frac{1}{2(A-1)T_A(b) \,
  f(x,Q^2)} \, \left\{ 1 \,-\, \exp\left[ -2(A-1)T_A(b) \, f(x,Q^2)
  \right] \right\} \;.
\eeq

\subsection{Results for gluon shadowing and validity of the model}
Summing up, in the framework of the Glauber-Gribov model the
total $\vphot A$ cross section can be calculated in a
straightforward way provided the
total $\vphot N$ cross section and the differential cross section
for diffractive production are known. In what follows, calculations
are made with the Schwimmer unitarization in Eq.~(\ref{eq:sch})
employing both FIT A and FIT B for the gluon density of the Pomeron.

\begin{figure}
  \centering
  \includegraphics[width=0.4\textwidth,height=8.cm]{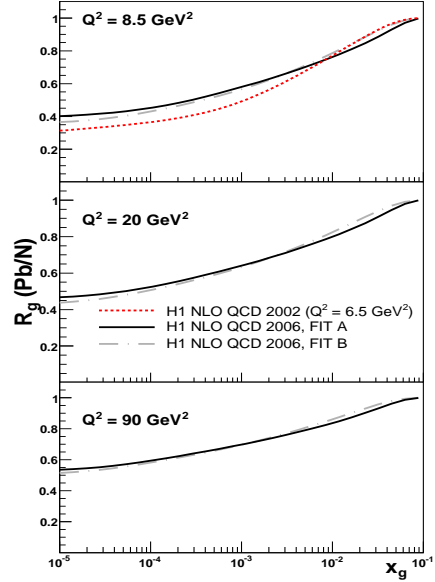}
  \caption{Gluon shadowing for Pb (lead) for different virtualities,
    $Q^2$.}
  \label{fig:ShadowingPb}
\end{figure}

The resulting gluon shadowing of the structure function of lead (Pb)
is depicted in Fig.~\ref{fig:ShadowingPb} for different virtualities,
$Q^2$. The shadowing is quite strong for $x < 10^{-3}$. We note that
the QCD evolution of the main term and 
rescattering terms are effectively treated separately in our approach,
and therefore the shadowing correction has a slow, logarithmic
dependence on $Q^2$ \cite{Brodsky02}. 
Details and further results
of the calculations can be found in \cite{Tywoniuk07}.

\begin{figure}
  \centering
  \includegraphics[width=0.7\textwidth]{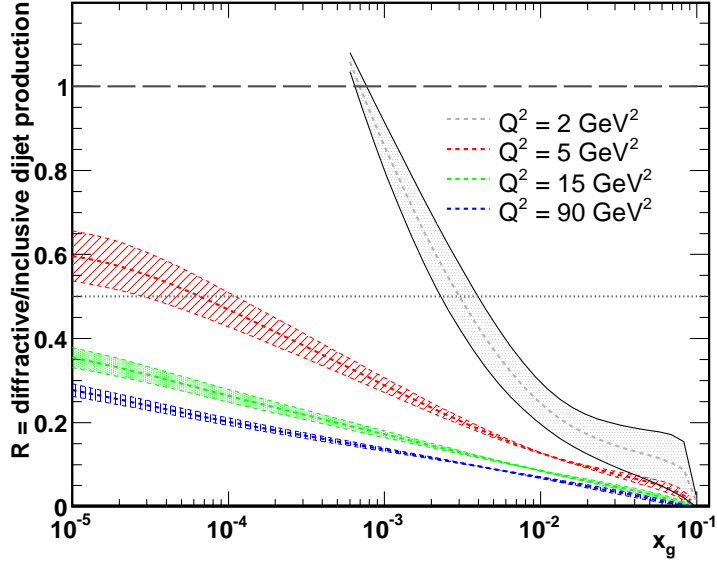}
  \caption{Pumplin ratio for different virtualities, $Q^2$. The shaded
    area denotes the uncertainty of diffractive gluon distribution
    function from HERA \cite{H1new1,H1new2}.}
  \label{fig:PumplinBound}
\end{figure}

Equation~(\ref{eq:sch})
does not take into account anti-shadowing effects which may play an
important role for $x \gsim 0.1$. 
The model for diffractive production by virtual photon described
above
is applicable for intermediate $Q^2 > 1-2
\mbox{ GeV}^2$ \cite{H1new1,H1new2} as the
parameterization of H1 data 
leads to a violation of unitarity for low $Q^2$ and $x \rightarrow 0$ and should be
modified at very low x. This is clearly seen in
Fig.~\ref{fig:PumplinBound} where we have plotted the so-called
Pumplin ratio \cite{Kaidalov03}
\beq
\label{eq:pump}
R \;=\; \frac{\int_x^{x_\Pom^{max}} \mbox{d}x_\Pom B(x_\Pom) \, f_\Pom
  (x_\Pom) \, \beta g^\D(\beta,Q^2)}{G_N (x,Q^2)} \;,
\eeq
which is in fact the ratio of diffractive and inclusive dijet
production. Unitarity is violated when $R \geq 1/2$. Calculations show
that this takes place at low $Q^2$ and low $x$. The shaded areas in
Fig.~\ref{fig:PumplinBound} represent the uncertainty in the diffractive
gluon distributions from FIT A and FIT B.
We conclude that our calculations are reliable in the
region $x > 10^{-4}$ which is relevant for RHIC and most experiments
at LHC, and should be taken with care for $x<10^{-4}$ at low $Q^2$. 

\section{Light hadron production at SPS and RHIC}
\begin{figure}
  \centering
  \includegraphics[width=0.6\textwidth]{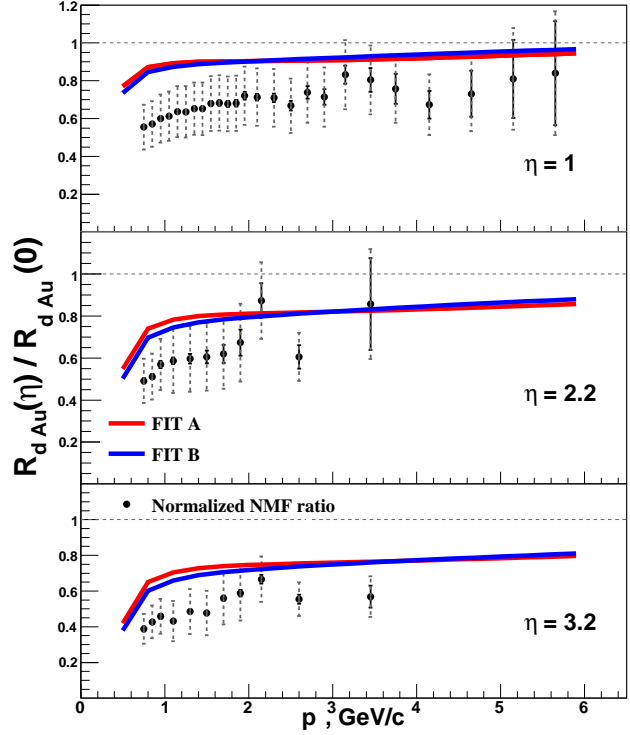}
  \caption{Ratio of forward and mid-rapidity nuclear modification
    factor. Gluon shadowing contribution to the suppression in $\eta$
    is depicted for two fits of diffractive gluon density. Data is
    taken from \cite{Arsene04}.}
  \label{fig:NMFratio}
\end{figure}

The model of gluon shadowing is now employed to study particle
production in dAu collisions at RHIC energy $\sqrt{s} = 200$ GeV.
There has been observed an increasing suppression of the nuclear
modification factor (NMF) 
\beq 
\label{eq:nmf} 
R_{d Au} (p_\bot,\eta) \;=\; \frac{1}{\left< N_{coll} \right>} \,
\frac{\mbox{d}^2N^{d Au}\, / \,\mbox{d} p_\bot \mbox{d} \eta}{\mbox{d}^2
  N^{pp}_{inel} \,/\,\mbox{d}p_\bot \mbox{d} \eta} 
\eeq 
with increasing pseudorapidity for charged particles
\cite{Arsene04}. We will not consider the effect of
$p_\bot$-broadening, or Cronin effect \cite{Cronin75}, at the moment,
but rather study the $\eta$ dependence of suppression. We assume, that
in the ratio
of forward to mid-rapidity nuclear modification factor
\beq
\label{eq:Rratio}
\tilde{R} \;=\; \frac{R_{d Au} (p_\bot,\eta)}{R^{norm}_{d Au} (p_\bot, 0)}
\eeq
the Cronin effect is cancelled out (at this energy the effect
is $<15$\%).
In Eq.~(\ref{eq:Rratio}),
$R_{dAu}^{norm}$ is the nuclear modification factor at mid-rapidity
divided by our calculations of gluon shadowing. The relation of
kinematical variables is given by the standard formula 
\beq
\label{eq:onejet}
x = \frac{c p_\bot}{\sqrt{s}} \, e^{-\eta} \;,
\eeq
where $c \sim 3$. 
In Fig.~\ref{fig:NMFratio} we see that gluon shadowing contribute to
the suppression of the nuclear modification factor at forward
rapidities, but is not sufficiently strong to describe the data
quantitatively.
Equation~(\ref{eq:onejet}) describes {\it de facto} mono-jet
production. The mean Bjorken x involved in particle production
calculated within perturbative QCD, which describes $2 \rightarrow 2$
particle collisions or so-called two-jet kinematics, is almost two
orders of magnitude larger \cite{Guzey04}. In this case, the effect of
shadowing on the suppression in Fig.~\ref{fig:NMFratio} would be
strongly reduced.
\par
So far we have neglected a very important part of the nuclear suppression
mechanism, namely the conservation of energy-momentum. This mechanism
is responsible for the observed shape of $\alpha(x_F)$ in the forward
region as seen in Fig.~\ref{fig:AlphaExperiment}. Energy-momentum
conservation affects the AGK cutting rules at finite energies
\cite{Capella76} and results in an additional suppression factor
\cite{Kopeliovich05,Boreskov93} corresponding to the substitution
\beq
\label{eq:enmom}
T_A(b) \;\longrightarrow\; T_A(b) \, \exp\left\{-\sigma^{eff}_{hN}(x_+,p_\bot) \,A
  T_A(b) \right\} \,
\eeq
in the numerator of Eq.~(\ref{eq:sch}). The shape of
$\sigma^{eff}_{hN}$ was formulated to describe data on low-$p_\bot$
nuclear modification factor for pPb collisions at SPS energy $\sqrt{s}
= 17.3$ GeV \cite{Boimska04}, namely
\beq
\label{eq:sigmaeff}
\sigma^{eff}_{hN} \left(x_+,p_\bot \right) \;=\; \frac{\sigma_0 \,
  x_+}{p_\bot^2 \,+\, p_0^2} \;,
\eeq
and $x_+ = 0.5\,(\sqrt{x_F^2 + 4 m_\bot^2/s} + x_F)$ where $m_\bot$ is
the transverse mass of
the produced particle. Once the two free
parameters, $\sigma_0$ and $p_0$, are fitted to data at $x_F = 0$,
Eq.~(\ref{eq:enmom}) describe the SPS nuclear modification factor for
$0 < x_F
\leq 0.4$. The same parameter values are also taken in the
calculations of the suppression at RHIC.\par
\begin{figure}
  \centering
  \includegraphics[width=0.7\textwidth]{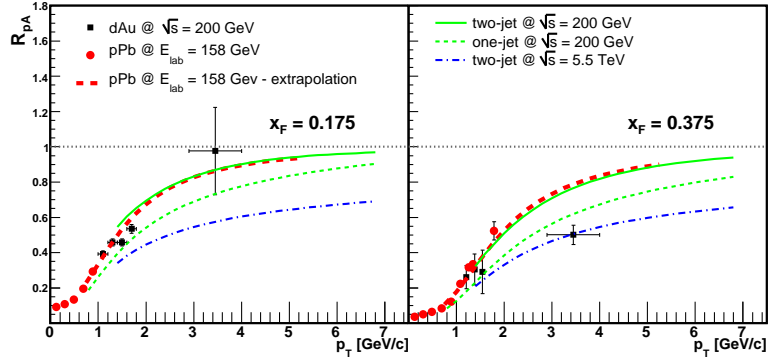}
  \caption{Comparison of nuclear suppression from pPb collisions at $\sqrt{s} =
  17.3$ GeV (red points) \cite{Boimska04}, and dAu collisions at $\sqrt{s} =
  200$ GeV (black points) \cite{Arsene04,Barnby04}, for two values of
  fixed $x_F$. Red curve is
  extrapolation of low-energy suppression to high $p_\bot$, green
  curve denotes the combined effect of shadowing and energy-momentum
  conservation at RHIC (solid: one-jet
  kinematics, dashed: two-jet kinematics). We present also predictions
  for pPb collisions at LHC energy (dash-dotted line).}
  \label{fig:xF}
\end{figure}

A comparison of SPS \cite{Boimska04} (red circles) and RHIC
\cite{Arsene04,Barnby04} (black squares) data on
nuclear suppression in pA collisions at $x_F = 0.175$ and $x_F =
0.375$ together with 
calculations of the combined effect of energy-momentum conservation
and gluon shadowing is presented in Fig.~\ref{fig:xF}. The red dotted
curve is an extrapolation of low-energy data to higher $p_\bot$ using
Eq.~(\ref{eq:enmom}). The green dashed curve show calculations using
Eq.~(\ref{eq:onejet}) for one-jet kinematics. We have also calculated
nuclear effects at RHIC using 
the mean Bjorken $x$ of a parton in the nucleus calculated within
pQCD (solid green curve in Fig.~\ref{fig:xF}). Predictions for LHC are
given as well (dash-dotted curves).\par
The most striking fact is that data at both energies seem to
overlap. This indicates that a common mechanism dominate
the suppression
at forward rapidities at both top SPS and top RHIC energies. Moreover,
since $\sigma^{eff}_{hN} (x_+ = 0)$ decrease with energy, RHIC data
gives room 
for an additional small shadowing contribution. Unfortunately, our
model give no information about the kinematics of particle production
at low and moderate $p_\bot$,
and so the 'one-jet' and 'two-jet' curves indicate the total
uncertainty of the model. At high-$p_\bot$, 'two-jet' kinematics is
theoretically more justified.

\section{Nuclear effects in heavy quarkonium production}
Production of heavy state, such as Drell-Yan and heavy-flavor, give
additional information on the energy dependence of 
nuclear suppression mechanisms. Since leptons interact very weakly with
the nuclear medium, Drell-Yan production holds information about the
initial state effects in pA collisions, e.~g. shadowing. On the other
hand, hidden heavy-flavor is believed to interact quite strongly with
the surrounding medium, either partonic or hadronic.
Recently, the substantial decrease of the nuclear absorption in 
$J/\psi$ production in hA collisions between SPS
energies, with $\sigma_{abs} \sim 4$ mb \cite{Ramello03}, and RHIC
energies, with $\sigma_{abs} \sim 1-2$ mb \cite{Adler03}, has
attracted a lot of attention as it is in contradiction with the
expectations of several theoretical models
\cite{Kopeliovich01,Braun98}. In what follows, we shall show
that the apparent observation of the reduction of $\sigma_{abs}$ can
be interpreted as a signal of the onset of coherent scattering
for heavy state production.

For heavy states the mass of the heavy system, $M_\QQ$,
introduces a new critical energy scale
\beq
\label{eq:scale}
s_M \;=\; \frac{M_\QQ^2}{x_+} \, \frac{R_A m_N}{\sqrt{3}} \;,
\eeq
It was shown in Ref.~\cite{Boreskov93} that AGK cutting rules are changed
at $s = s_M$. At energies below $s_M$ longitudinally ordered
rescattering of the heavy system takes place and the produced heavy
system is subject to nuclear absorption. In this energy interval
Drell-Yan production does not experience nuclear
suppression. At $s > s_M$ the heavy state in the projectile, which
also includes light degrees of freedom, scatters coherently off
nucleons of a nucleus and the conventional treatment of nuclear
absorption is not adequate. For $x_F$ close to zero (central rapidity
region) values of $s_M$ for $J/\psi$ and $\Upsilon$ belong to the RHIC
energy region. In this kinematical region the effects of shadowing of
nuclear partons become important. A similar approach for the description of
$J/\psi$-suppression in dAu collisions at RHIC has been considered in
Ref.~\cite{Capella06}. For $x_F \sim 1$ the high-energy regime corresponds
to an interaction with a nucleus of the fluctuation of a projectile,
containing heavy quarks (of the type of ``intrinsic charm'' mechanism
of Refs.~\cite{Vogt00,Brodsky80}).

Thus at energies $s < s_M$ the inclusive cross section for production
of particle $a$ in a hA collision is given by
\beq
\label{eq:Abs}
E\,\frac{\mbox{d}^3 \sigma^a_{hA}}{\mbox{d}^3 p}
\left( x_+ \right) \;=\;  E\,\frac{\mbox{d}^3
  \sigma^a_{hN}}{\mbox{d}^3 p} \left( x_+ \right) \, \int
\mbox{d}^2b \, \frac{1 - 
  \exp \left[- \xi(x_+) \sigma_\QQ A T_A (b) \right]}{\xi (x_+)
  \sigma_\QQ}\, R_g^{Sch} \left(b,x_+,p_\bot \right) \;,
\eeq
where $R_g^{Sch}$ is the $b$-dependent shadowing factor calculated in
Eq.~(\ref{eq:bshad}) and
$\xi (x_+) = (1-\epsilon) + \epsilon x^\gamma_+$ determines the
$x_+$ dependence of absorption. The suppression is
concentrated at much  
higher $x_F$ for $Q\bar{Q}$ production than for the light hadrons because
of the large mass of the $Q\bar{Q}$ system, e.~g. $\gamma = 2$ for
$J/\psi$ and $\gamma \sim 3$ for $\Upsilon$ production.
Equation~(\ref{eq:Abs}) gives a good
description of experimental data on charmonium production in pA
collisions at $E_{LAB} \lsim $ 800 GeV/c with $\sigma_\QQ = 20$ mb and
$\epsilon = 0.75$ \cite{Boreskov93}. This corresponds to an absorption cross section of
$\sigma_{abs} = 5$ mb. Note that $\sigma_\QQ$ is rather large,
indicating that the $c\bar{c}$ pair is produced in the color octet
state rather than in the colorless state. It can
also be viewed as a $D\bar{D}$ ($D^*\bar{D}^*$) system.

Equation (\ref{eq:Abs}) is not valid at
asymptotic energies as the assumption of longitudinal
ordering leading to it is only valid at $s < s_M$. For
energies higher than $s_M$ the
expression will change due to the correct treatment of coherence
effects \cite{Braun98}
\beq
\label{eq:transition}
\frac{1 \,-\, \exp \left[-\xi (x_+)\sigma_\QQ A T_A(b) \right]}{\xi
  (x_+) \sigma_\QQ} \;\longrightarrow \; A T_A (b) \, \exp \left(-
  \sigma^{eff}_{\QQ} (x_F) \, A T_A(b)/2 \right) \;,
\eeq
which is similar to the energy-momentum conservation effect
for light quarks. In the model of Ref.~\cite{Braun98},
$\sigma^{eff}_\QQ$ was found to be 
the $Q\bar{Q}-N$ total cross section. 

We would like to argue that this leads to an incorrect behaviour at
high energies, and propose an alternative procedure. Considering
non-enhanced, Glauber-type diagrams 
the effective cross section $\sigma_\QQ^{eff}$ should be proportional
to $x_+^\gamma$,
thus satisfying AGK cancellation in this limit. 
It was shown in Ref.~\cite{Boreskov93} that at $x_F \sim 1$ the second
rescattering in the low and high energy limits should coincide. This
means that $\sigma^{eff}_\QQ \approx \epsilon x_+^\gamma
\sigma_\QQ$. Experiment on $J/\psi$ production in dAu
collisions at RHIC \cite{Adler03} was performed in the central
rapidity region, where $x_+ \sim 0.025-0.05$ and $\sigma^{eff}_\QQ$ is therefore
very small. This favors the pure nuclear shadowing scenario, which
means that we do not include contributions from non-enhanced,
Glauber-type diagrams.
\par
\begin{figure}
  \centering
  \includegraphics[width=0.5\textwidth,height=8.5cm]{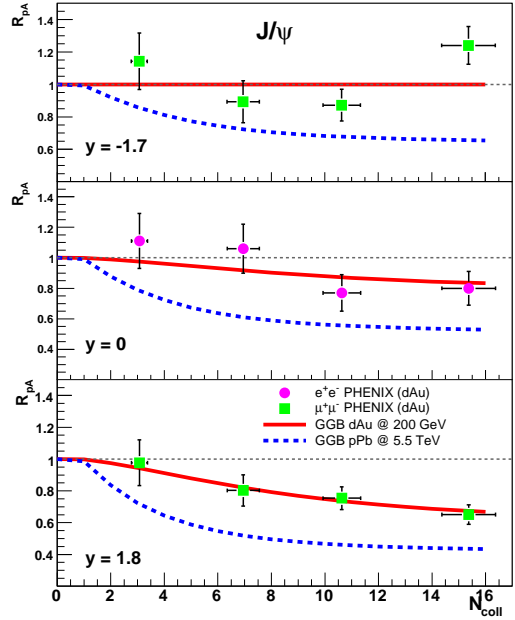}
  \caption{Centrality dependence of the nuclear modification factor in
    dAu (pPb) collisions at $\sqrt{s} = $ 200 GeV (5.5 TeV) for $J/\psi$ at
    different rapidities. Data are taken from \cite{Adler03}.}
  \label{fig:JpsiCentrality}
\end{figure}
\begin{figure}
  \centering
  \includegraphics[width=0.7\textwidth]{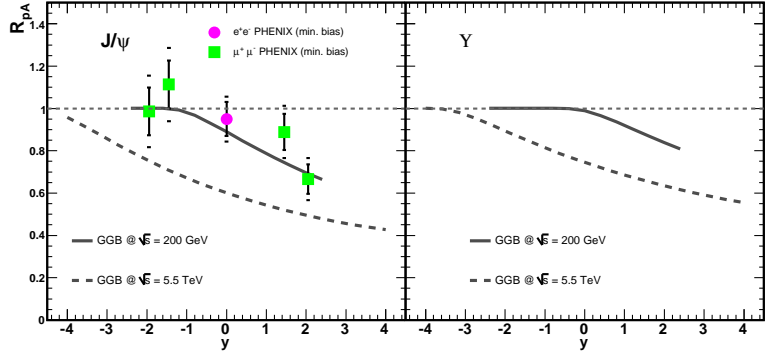}
  \caption{Rapidity dependence of the nuclear modification factor in
    min. bias
    dAu collisions (pPb) at $\sqrt{s} = $ 200 GeV (5.5 TeV) for $J/\psi$ and
    $\Upsilon$ at different rapidities. Data are taken from \cite{Adler03}.}
  \label{fig:JpsiUpsilonRapidity}
\end{figure}
Results of our calculations for $J/\psi$ and $\Upsilon$ production in
dAu collisions at RHIC are
shown in Figs.~\ref{fig:JpsiCentrality} and
\ref{fig:JpsiUpsilonRapidity} together with data from \cite{Adler03}
(solid curves). Both the centrality and the rapidity dependence of the
data is well reproduced. We predict a stronger shadowing effect in pPb
collisions at LHC energy $\sqrt{s} = $ 5.5 TeV, which is given by the
dashed curves in Figs.~\ref{fig:JpsiCentrality} and
\ref{fig:JpsiUpsilonRapidity}. This effect should also be taken into
account when modeling nucleus-nucleus collisions at high energy.

\section{Conclusions}
We have calculated gluon shadowing in the Glauber-Gribov model using
latest parameterization of diffractive gluon distribution from HERA. A
strong shadowing effect is found. Our model is applicable for $Q^2> 2$
GeV$^2$ and $10^-5 < x < 0.1$, well suited for analysis of moderate and
high-$p_\bot$ particle as well as heavy-flavor production at
high-energy experiments.

Particle production at RHIC is dominated by coherent production both
for light and heavy particles at low and moderate $p_\bot$. This is
most clearly observed at 
mid-rapidity where $\alpha(x_F=0)$ is below but close to
unity for both charged particles and $J/\psi$ - the small suppression
is solely due to gluon shadowing. Shadowing effects are stronger for
light than for heavy particles as expected.

Both centrality and rapidity dependence of $J/\psi$ production in dAu
collisions at RHIC have been described with our model.
Even so, gluon shadowing alone cannot explain the $\eta$ dependent
suppression of light particles measured at RHIC.
At forward rapidities energy-momentum conservation comes into play and
contributes strongly to the observed suppression. This happens more rapidly
for light particle production, for $J/\psi$ and
$\Upsilon$ this effect is shifted to larger values of $x_F$ due to the
large mass. The combined effect of energy-momentum conservation and
gluon shadowing shows good agreement with SPS and RHIC data.

The discussion of mid-rapidity nuclear modification factor is out of
scope of this paper. A detailed calculation with
the above mentioned effects and Cronin effect should be performed at
all rapidities to draw quantitative conclusions from experimental
data. Gluon shadowing by itself can also be checked against dilepton
or direct photon data, and also in ultra-peripheral heavy-ion
collisions.

\acknowledgments
The authors would like to thank N.~Armesto, K.~Boreskov, V.~Guzey,
D.~R\"ohrich and M.~Strikman for 
interesting discussions, and B.~Boimska for providing experimental
data.
This work was supported by the Norwegian
Research Council (NFR) under contract No.~166727/V30, QUOTA-program,
RFBF-06-02-17912, RFBF-06-02-72041-MNTI, INTAS 05-103-7515, grant of
leading scientific schools 845.2006.2 and support of Federal
Agency on Atomic Energy of Russia.

\end{document}